# Does Crypto Kill?
## Relationship Between Electricity Consumption Carbon Footprints and Bitcoin Transactions

Altanai Bisht[1], Arielle Wilson[1], Zachary Jeffreys[1], Shadrokh Samavi[1,2]
[1]Computer Science Department, Seattle University, Seattle, 98122 USA
[2]Elect. & Comp. Engineering, McMaster University, L8S 4L8, Canada

*Abstract*— Cryptocurrencies are gaining more popularity due to their security, making counterfeits impossible. However, these digital currencies have been criticized for creating a large carbon footprint due to their algorithmic complexity and decentralized system design for proof of work and mining. We hypothesize that the carbon footprint of cryptocurrency transactions has a higher dependency on carbon-rich fuel sources than green or renewable fuel sources. We provide a machine learning framework to model such transactions and correlate them with the electricity generation patterns to estimate and analyze their carbon cost.

## 1- INTRODUCTION

Climate change is a universal problem, and a large part of this is the carbon footprint from electricity consumed by digital transactions. These transactions range from simple web page requests to high-definition video calls. Digital currencies such as non-fungible token (NFT) exchanges and cryptocurrency transactions on a decentralized distributed blockchain involve mining, which has been criticized for having a high carbon footprint due to the high computing power used. Carbon emissions for these cryptocurrency transactions vary according to fuel sources that make up the fuel mix of a grid at different times of the day, week, month, or even season. The various energy sources have different carbon emission lifecycles. The lifecycle affects the carbon footprint of the transactions. Our model will cluster cryptocurrency transactions based on the electricity generation fuel mix for carbon footprint tracking.

We open-source our model and associated research to facilitate potential future works in this field. We concentrate on Bitcoin transactions, but our model is usable for other digital currencies and blockchain frameworks. The main goal of our model is to determine if cryptocurrency generation is economically beneficial or will it doom us all with a heavy carbon footprint.

The structure of the paper is as follows: in section 2, we will explain the preprocessing of the data. Section 3 details our experiments. In section 4, we show the relationship between carbon footprint and the volume of transactions. Finally, section 5 lists the findings of this research, and concluding remarks are presented in section 6.

## 2- PREPROCESSING

We train our model for blockchain analysis on blockchain-based cryptocurrency Bitcoin with electricity generation from California ISO. The aim is to analyze whether BTV (Bitcoin Transaction volume) has a greater reliance on carbon-rich fuel sources or green renewable fuel sources. We focus our study on California as the hub for most transactions in Bitcoin. Studies depicting the hotspots of Bitcoin transactions show large transactions in California [7]. The experiments use many datasets from various sources, such as:

A. Electricity Generation by U.S. Energy Information Administration (EIA)[1]
B. Bitcoin Blockchain Historical Data in Kaggle[2]
C. Carbon life cycle emission by IPCC[9]
D. Average Temperature by National Centers for Environmental Information (NCDC)[10]
E. Bitcoin's carbon footprint for classification validation from the Cambridge study[3] and EIA Co2 emission per Mw[4]

Although our dataset does not contain any missing data, it has some inconsistencies in timestamp sampling. Therefore, to ensure our dataset's accuracy, completeness, and consistency, we extrapolate, resample, normalize, transform and scale the dataset to fit into the model.

*A. Electricity generation in CAISO*

The fuel mix of a region changes with daily temperature fluctuations, weather changes, seasonal changes, daylight savings, and crude oil prices, among many other factors. We use the California ISO grid's generation fuel mix to model the carbon footprint of cryptocurrency transactions. Figure 1 shows the fuel mix of the California region's electricity mix in July of 2021.

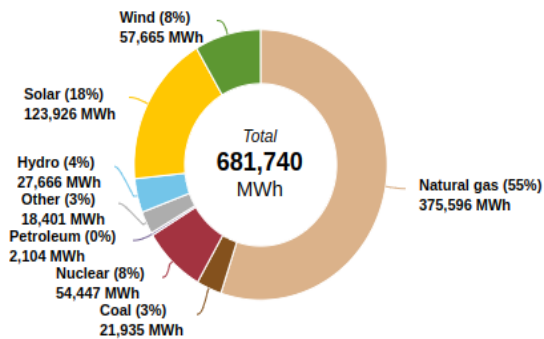

Fig.1 CAISO (California Independent System Operator) grid fuel mix in July 2021

*B. Bitcoin transaction Volume*

Cryptocurrency Bitcoin's Transactions Volume (BTV) data, as shown in Fig. 2, has many outliers. These outliers indicate the non-deterministic and volatile behavior of cryptocurrency transactions. To use the BTV for modeling fuel usage, we scale the range using a base-2 logarithm and further normalize the dataset.

## 3- EXPERIMENTATION

When a block is created in a blockchain, a computer solves a hash function to find a nonce number that only happens once. There is a 1 in 1.36 million chance [6] of solving the hash function, requiring many brute force iterations. The heavy processing servers that mine blocks heat up, drawing much more power from the grid than a computer would, contributing to global warming. On top of this, blockchain miners need to keep these servers cool. The cooling mechanisms draw power, adding even more to the high carbon footprint of each transaction.

For modeling the high or low carbon footprint of a transaction, we gather the fuel mix of every hour of electricity generation. We used these features:

1. Carbon-rich fuel (such as petroleum, natural gas, and coal) generation in Mega Watts (M.W.)
2. Green-renewable fuel(such as nuclear, solar, hydro, and wind) generation in Mega Watts (M.W.)
3. Bitcoin's transaction volume (BTV)
4. The average temperature of each day in California for the calendar year under calculation (TAVG)
5. The carbon footprint from each fuel

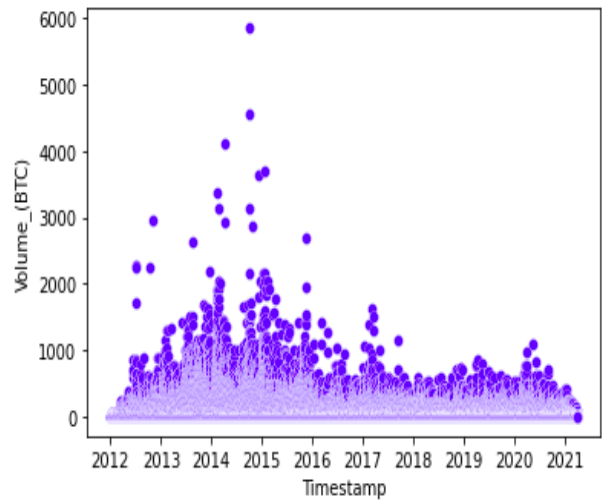

Fig.2 Bitcoin transactions from 2012 to present

*A. Hourly Data Analysis*

We exclusively considered all Bitcoin transactions in 2021 and the green or carbon-rich electricity generation fuel mix from the grid, as shown in Fig.3(a). The cluster formation from various methodologies (agglomerative using silhouette scores, K-Means with knee-locator, MiniBatch-K-Means, and Mean-Shift) suggests a higher contribution by carbon-rich fuel sources than green renewable sources, also depicted in Fig. 3(b).

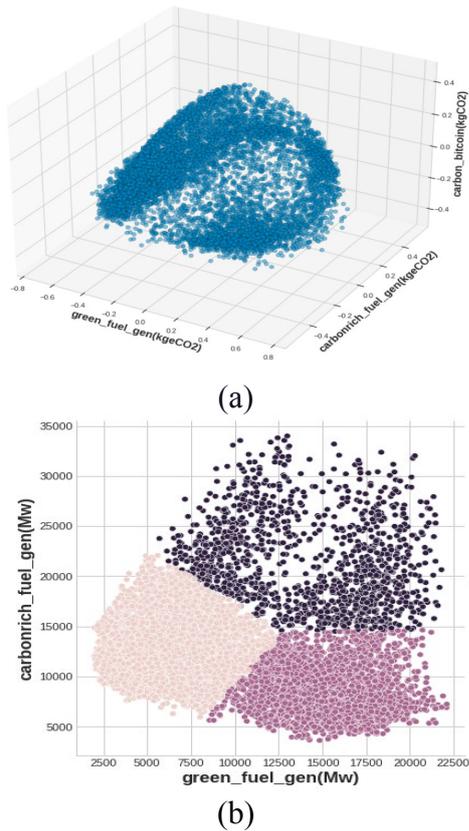

Fig. 3 (a) Unclustered distribution, (b) a 2-dimensional distribution of BTV corresponding to carbon-rich fuel and green renewable fuel usage.

**Optimal number of clusters**: Depicted in Fig. 4, the results from the elbow locator method to find an optimal number of clusters. In this method, different numbers of clusters are tested, and the sum squared error (SSE) is calculated. Our experiments show that three is a suitable number of clusters.

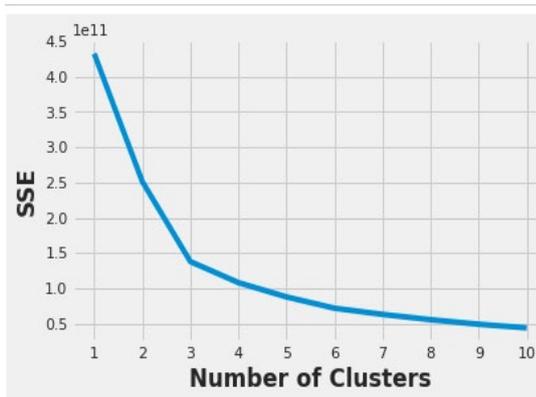

Fig.4 Knee locator estimates the optimal number of clusters for Bitcoin transactions to be three.

Accordingly, minibatch K-means is parameterized to create three clusters, as shown in Fig.5.

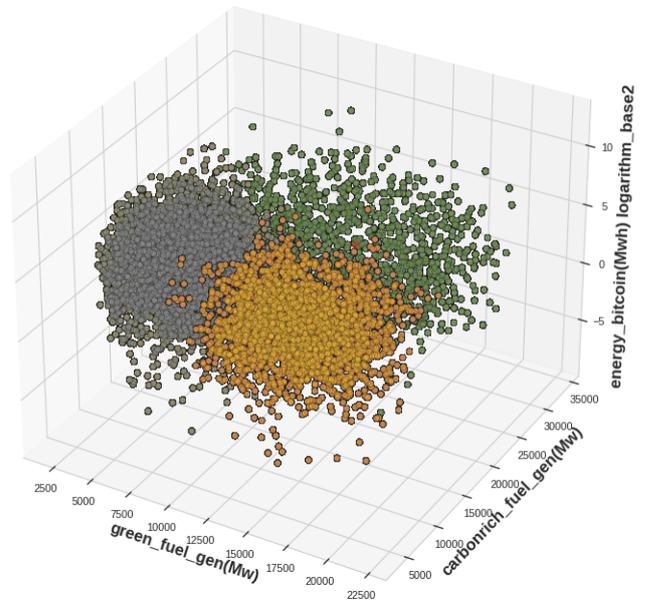

Fig.5 Hourly Bitcoin transactions distributed across three clusters using supervised Minibatch-K-means learning method

The clusters show electricity consumption of Bitcoin transactions from carbon-rich and green-renewable sources.

**Clustering:** Mean-shift clustering applied to the transformed and scaled BTV data results in many suggestive clusters, as shown in Fig.6.

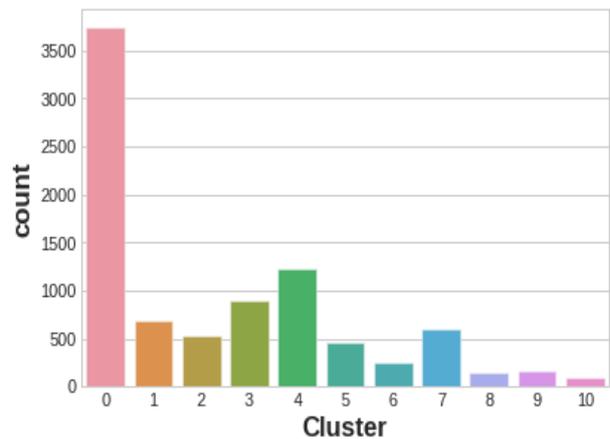

Fig.6. Distribution of Bitcoin transactions using Mean-shift clustering.

However, as Bitcoin transaction's electricity consumption is clustered into green-renewable and carbon-rich fuel usage, as shown in Fig.7. Therefore, the categorization is based on carbon-rich fuel usage and further sub-categorization into green fuel usage within a category of carbon-rich fuel usage.

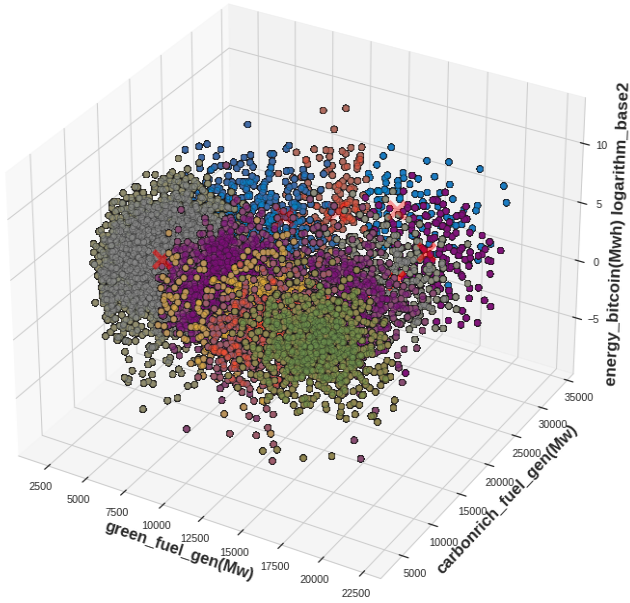

Fig.7 Hourly Bitcoin transactions distributed across many clusters using the Mean-Shift learning method

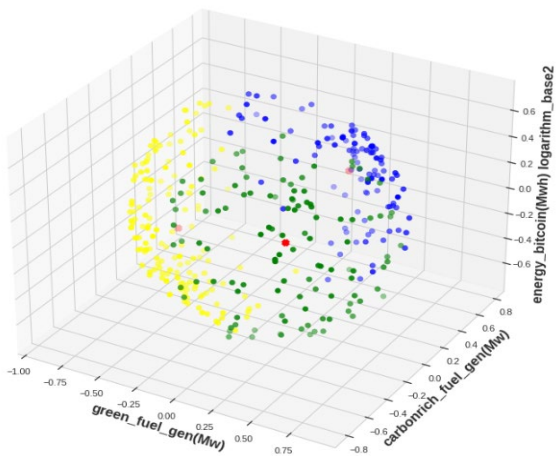

(a)

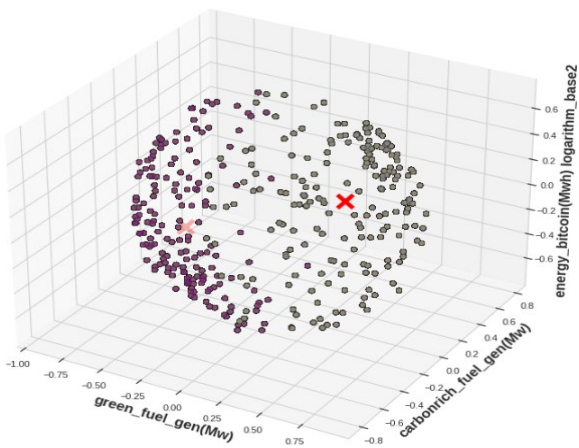

(b)

Fig.8 Daily Bitcoin transactions cluster formation in 3D scatter plots with centroids using (a) Supervised and (b) unsupervised (right) learning methods.

### B. Daily Data Analysis of Bitcoin Transactions Fuel usage

We perform resampling on the hourly electricity consumption of BTV to daily samples. Clustering divides the BTV into

1. Low carbon-rich fuel, high green- renewable energy fuel usage and
2. High carbon-rich fuel, low green energy fuel usage.

The clustering of the BTV is shown in Fig. 8. The transaction that can be identified as "1. Low carbon-rich fuel, high green- renewable energy fuel usage" is significantly lower than "2. High carbon-rich fuel, low green energy fuel usage".

### C. Weekly Data Analysis of Bitcoin Transactions Fuel usage

Depicted in Fig.9 is the trend of Bitcoin consumption and electricity generation by green-renewable and carbon-rich fuel sources. In addition, the trend outlines how Bitcoin consumption is more aligned with electricity generation from carbon-rich fuel sources.

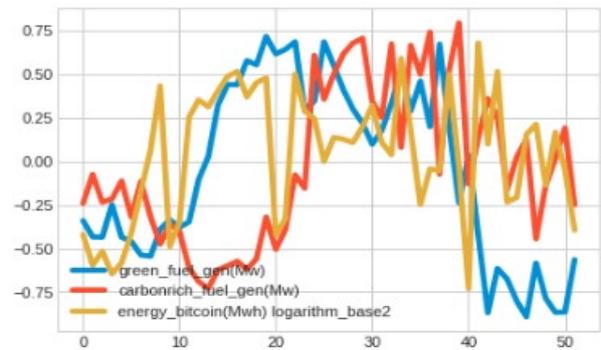

Fig.9 BTV trend of increasing weekly sampled Bitcoin transactions and increasing the electricity generation from both groups of fuel sources.

A supervised and unsupervised cluster analysis shows a similar trend of grouping weekly Bitcoin in Fig.10(a) and 10(b) respectively. In both learning methods the location of the larger clusters and centroids is in high carbon-rich fuel usage as compared to renewable-green fuel usage.

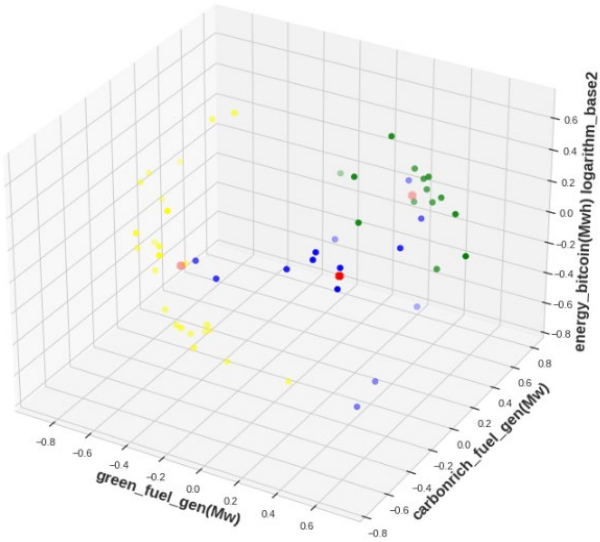

(a)

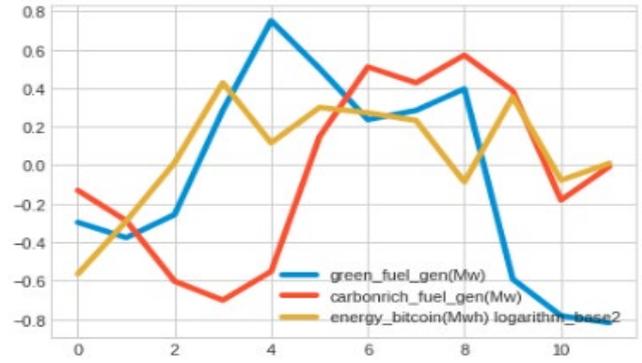

Fig. 11 BTV trend of increasing monthly sampled Bitcoin transactions and increasing the electricity generation from both groups of fuel sources.

The monthly analysis also follows the same trend outlined by prior experiments showing more relevance of carbon-rich fuel as energy for BTV is consumed more from carbon-rich fuel sources than green-renewable ones between Oct-Dec.

Fig.12 shows that monthly sampled Bitcoin transactions have high carbon-rich fuel usage while relatively low green-renewable fuel usage.

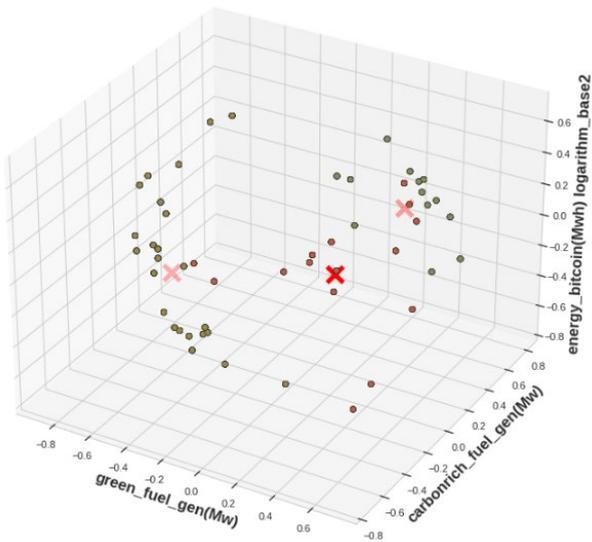

(b)

Fig.10 Weekly Bitcoin transactions cluster formation in 3D scatter plots with centroids using (a) Supervised and (b) unsupervised (right) learning methods

*D. Monthly Data Analysis of Fuel usage in Bitcoin Transactions*

The monthly sampled trend of BTV is shown in Fig.11. The trend shows that the months in the first part of the year (Jan-May) have a low carbon footprint of transactions than later months (Jun-Dec). Fig. 11 also shows how Bitcoin has approximately uniform transactions every month except for January in 2021, where there is a significant dip.

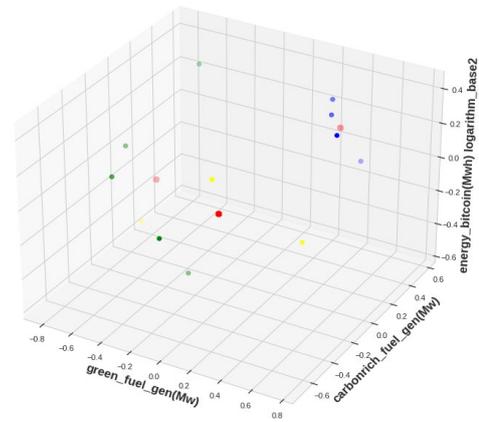

(a)

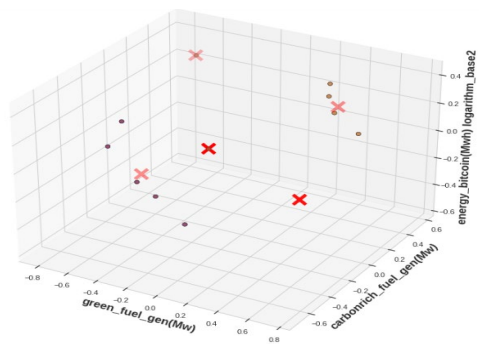

(b)

Fig.12 Monthly Bitcoin transactions cluster formation in 3D scatter plots with centroids using (a) Supervised and (b) unsupervised (right) learning methods

## E. Temperature based analysis of Bitcoin transactions

Figure 13 shows the variation in Average Temperature (TAVG) in the California region in 2021.

The correlation between TAVG and the total energy generated for 2021 in the California region is 0.738546. At the same time, the correlation between BTV and TAVG is 0.247355. Hence energy consumed by BTV is affected by both the temperature and the total energy generated. Interestingly with average temperature as one of the principal fields in the dataset, the Bitcoin transactional data clusters into majorly two sets

1. Normal temperature, low energy consumption transactions.
2. Hotter temperature, high energy consumption transactions.

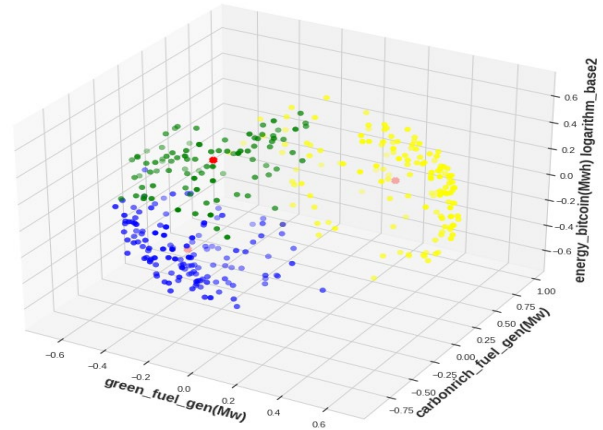

(a)

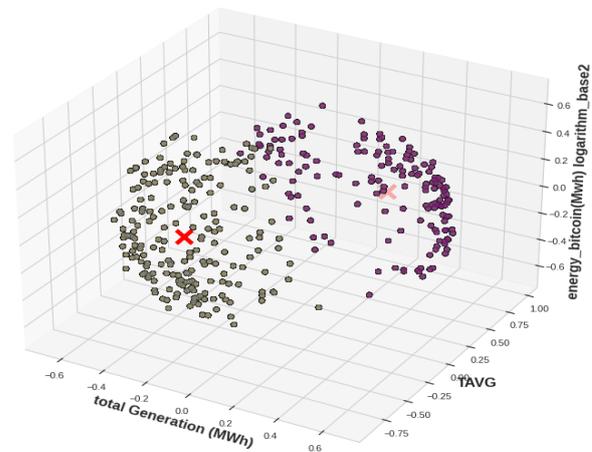

(b)

Fig.14 Supervised (a) and unsupervised (b) cluster formation in 3D scatter plots with centroids, according to the average temperature in the region.

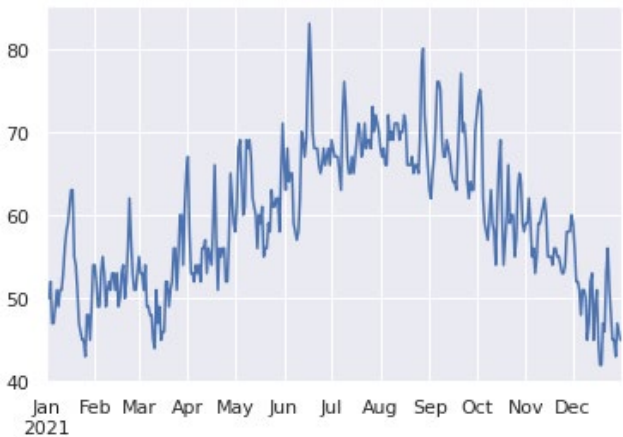

Fig. 13 Average temperature in the grid observation region (California) in 2021.

This clustering also impresses the notion that most cryptocurrency transactions occur in either a very hot or very cold spectrum of temperature, which is also depicted in both figures Fig.14(a) and 14(b).

Given a strong correlation between total generation and carbon-rich fuel supply, 0.876389, we know that carbon-rich fuel is used in peaking power demands in very low or high temperatures. Thus high total generation with extreme temperatures results in higher carbon-rich fuel usage by cryptocurrency transactions.

## 4- CARBON FOOTPRINT

We used LCA (Lifetime Carbon Analysis) from electricity generation as the features to develop the model. The parameters used are:

- Carbon-rich fuel generation's carbon emission (unit kgeCO2) in grid
- Green/ renewable fuel generation's carbon emission (unit kgeCO2) in grid
- Bitcoin's estimated carbon footprint based on the fuel mix consumed.

We know that cumulative carbon emission from any digital transaction increases over the years, which also applies to Bitcoin, as shown in Fig.15(a). The predicted carbon emission from carbon-rich and green-renewable fuel usage and their LCA values is shown in Fig. 15(b).

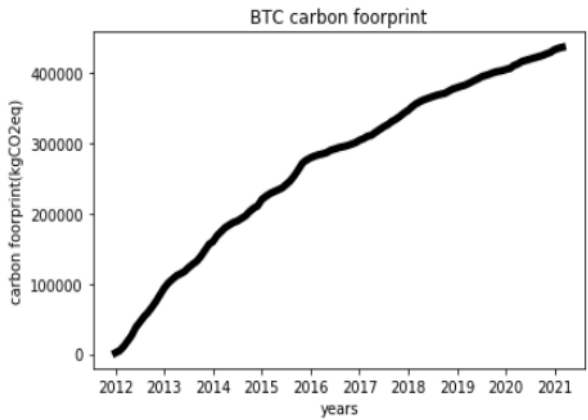

(a)

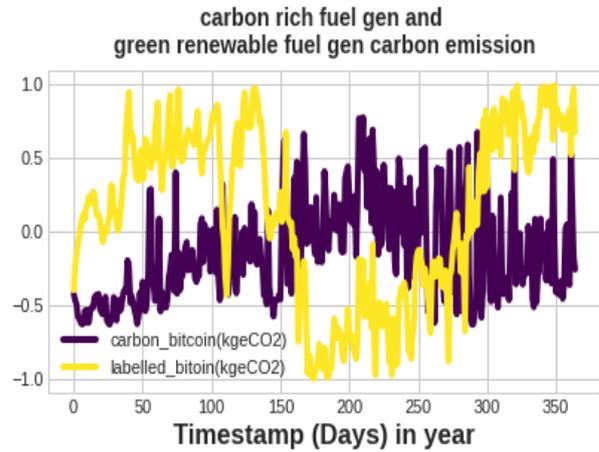

(a)

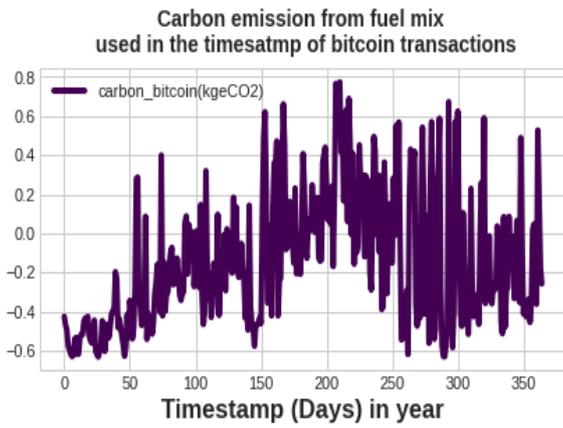

(b)

Fig. 15. (a) Increasing carbon footprint of Bitcoin transactions over the decade and (b) fluctuation in Bitcoin carbon emission in 2021.

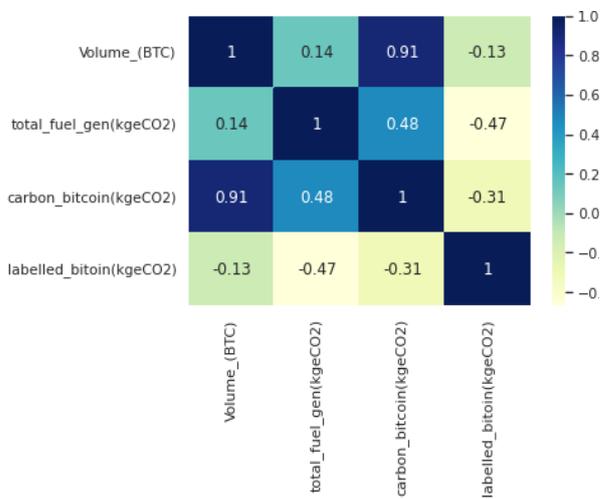

(b)

Fig. 16 (a) Carbon footprint of Bitcoin transactions from CBECI and fuel usage analysis. (b) Heatmap for bitcoin transaction volume correlation with other parameters as carbon emission.

To validate our Bitcoin transactions classification into classes based on high and low carbon emissions, we use Cambridge Bitcoin Electricity Consumption Index (CBECI) [8]. The following factors are considered: the network hash rate, Bitcoin issuance value, miner fee, Bitcoin market price, mining equipment efficiency, electricity cost, power usage effectiveness, and hash rate share.

However, our computed Bitcoin's carbon footprint does not align with Cambridge's CBECI, as shown in Fig.16(a) and 16(b). This inconsistency is because of a sudden dip in the CBECI energy consumption in the middle of the year. The mid-year dip does not correlate to the volume of Bitcoin transactions in the same period.

As Bitcoin transactions consume more carbon-rich fuel than green renewable fuel, higher carbon emissions increase if the transaction volume increases. Fig.17 shows a linear trend toward increasing carbon emission with increasing volume due to high reliance on carbon-based fuel sources.

## 5- FINDINGS

Critical findings using this model and analysis are

*1)* Most crypto transactions fall under the "High Carbon-rich, low green fuel usage" cluster. We showed that Bitcoin transactions rely more on carbon-based fuel sources than green renewable fuel sources from the electricity grid.

2)   Seasonal and temperature changes affect Bitcoin transactions. For example, when temperatures are on the far end of the spectrum, such as hot and cold average temperatures, the effects of the Bitcoin transactions are more pronounced.

3)   Consequently, the carbon footprint of the Bitcoin transactions has more contribution from carbon emissions produced by carbon-rich fuel sources than green or renewable sources, which is as expected since green and renewable sources have a low carbon footprint themselves.

4)   While low Bitcoin transaction volume has a lower range of carbon emissions, high volume traffic has a broader range of emissions which pertain to the timestamp of the transactions during a day since the fuel mix of the electricity grid varies across the day.

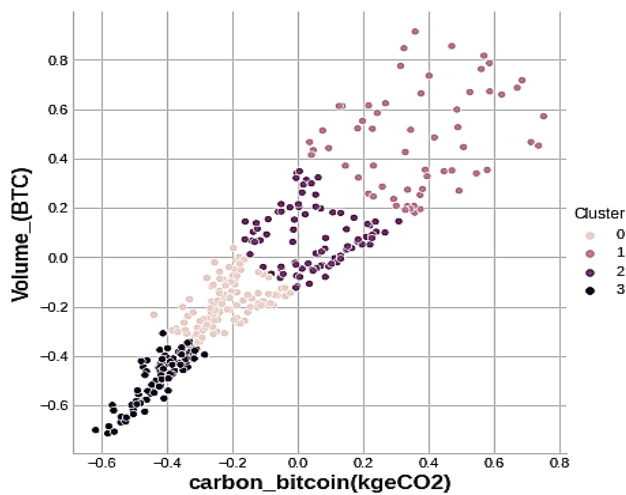

Fig. 17 Carbon footprint of Bitcoin transactions mapped against BTV

## 6- CONCLUSIONS

This research analyzed if cryptocurrencies such as Bitcoin harm our environment. Furthermore, we showed how the carbon footprint of cryptocurrency transactions could be obtained from the fuel mix in the grid of the electricity generator. Based on this research, we proved that the frequency and timestamp of transactions provide insight into the fuel sources consumed from the main power grid. The fuel consumption, in turn, can be used in the calculation of the carbon footprint of the cryptocurrency transactions by using lifetime carbon analysis (LCA) of the fuel sources.

We trained our model to classify and cluster these cryptocurrency transaction data into fuel source reliance. We found that most transactions align with carbon-rich fuel sources rather than green renewable sources. Thus, crypto, or at least Bitcoin, negatively impacts our environment.

This open-source academic project also provides a carbon footprint analysis framework for any I.P. transaction based on its trends and geographical electricity generation fuel mix

## ACKNOWLEDGMENT

All of the data in this research come from U.S. Energy Information Administration (EIA) [1].